# Thermal Hall effect in a van der Waals triangular magnet FeCl$_2$

Chunqiang Xu[1,2], Caitlin Carnahan[3], Heda Zhang[1], Milos Sretenovic[1], Pengpeng Zhang[1], Di Xiao[4,5], and Xianglin Ke[1]

[1]*Department of Physics and Astronomy, Michigan State University, East Lansing, Michigan 48824-2320, USA*

[2]*School of Physical Science and Technology, Ningbo University, Ningbo 315211, China*

[3]*Department of Physics, Carnegie Mellon University, Pittsburgh, Pennsylvania 15213, USA*

[4]*Department of Materials Science and Engineering, University of Washington, Seattle, Washington 98195, USA*

[5]*Department of Physics, University of Washington, Seattle, Washington 98195, USA*

Thermal transport is a pivotal probe for studying low-energy, charge-neutral quasi-particles in insulating magnets. In this Letter, we report an observation of large magneto-thermal conductivity and thermal Hall effect (THE) in a van der Waals antiferromagnet FeCl$_2$. The magneto-thermal conductivity reaches over ~700%, indicating strong magnon-phonon coupling. Furthermore, we find an appreciable thermal Hall signal which changes sign concurrently with the spin-flip transition from the antiferromagnetic state to the polarized ferromagnetic state. Our theoretical calculations suggest that, in addition to the Berry curvature induced at the anticrossing points of the hybridized magnon and acoustic phonon modes of FeCl$_2$, other mechanisms are needed to account for the magnitude of the observed THE.

Thermal transport is an important technique for studying low-energy excitations of quantum materials. For instance, the thermal Hall effect (THE), i.e., the generation of a transverse heat current carried by charge-neutral quasiparticles in the presence of a longitudinal temperature gradient, has been attracting more and more attention since the experimental observation of THE in the paramagnetic insulator $Tb_3Ga_5O_{12}$ reported by Strohm et al in 2005 [1]. In particular, the THE may be used to probe the characteristic features of emergent quantum phases. For instance, THE measurements provide pivotal evidence of topological magnons in magnetic insulators [2-4], chiral phonons in cuprate superconductors [5,6], and potential Majorana-fermions in Kitaev quantum spin liquid candidates [7-10].

In magnetic solids with strong magnetoelastic coupling, magnon-phonon coupling can modify the features of both magnetic excitations and lattice vibrations. For instance, magnons and phonons can scatter off one another, leading to the shortened lifetimes and the resulting suppression of heat conduction carried by these quasiparticles. Furthermore, magnon and phonon modes may hybridize, leading to the emergence of coherent quasiparticles known as magnon-polarons [11-13]. Compared to the band structures of non-interacting magnon and phonon modes which cross each other, a salient character of the magnon-phonon hybridized excitations is the appearance of anticrossing points leading to a gap opening at the intersections of magnon and phonon dispersions. Recently, it has been shown that these anticrossing points give rise to Berry curvature hotspots [4,14-18]. As a result, magnon-polarons may exhibit nontrivial topological character with the hybridized bands carrying non-zero Chern numbers, which are then anticipated to yield a THE [15-18]. Despite the increasing theoretical attention and potential applications in spin caloritronics and magnon spintronics [19], experimental realization of magnon-polarons-driven THE is very rare.

In this Letter, we report the observation of large magneto-thermal conductivity and THE in a van der Waals (vdW) antiferromagnetic insulator $FeCl_2$ hosting a triangular magnetic lattice. The magneto-thermal conductivity reaches over ~700%, indicating strong magnon-phonon coupling. The THE increases with magnetic field prior to the spin-flip transition from the antiferromagnetic state to the polarized ferromagnetic state, at which the THE exhibits a sign change followed by a decrease in magnitude upon further increasing the field. Our theoretical calculations suggest that magnon-polarons resulting from the hybridized magnon and acoustic phonon modes in $FeCl_2$ can be associated with the observed THE. The discrepancy in the magnitude of thermal Hall conductivity between the experimental data and the theoretical calculation indicates the possibility of other contributing mechanisms which call for future studies.

$FeCl_2$ crystalizes in the space group $R\bar{3}m$ (No. 166) at room temperature with lattice constants $a = b = 3.600(3)$ Å, $c = 17.539(2)$ Å and crystalline angles of $\alpha = \beta = 90°$, $\gamma = 120°$ [20]. As illustrated in Fig. 1(a,b), $FeCl_2$ is a vdW material with the iron atoms forming a triangle structure in the $ab$-plane. Previous neutron measurements showed that $FeCl_2$ is an A-type antiferromagnet below $T_N \sim 23.5$ K with neighboring ferromagnetic planes weakly coupled antiferromagnetically along the $c$-axis and the magnetic easy-axis along $c$ [21,22]. Owing to the strong spin-orbit coupling, its magnetic excitation exhibits a large single-ion anisotropy gap of ~ 2.2 meV, and its magnon dispersions can be nicely described using the spin wave theory on a two-dimensional ferromagnet with effective $S = 1$ moment [23]. Previous theoretical studies suggested that the magnon and phonon modes may hybridize near the zone center in this system [24], motivating us to investigate the thermal Hall effect. Detailed information regarding the material synthesis and experimental methods can be found in the Supplemental Materials [25] which also includes Ref. [26-28].

Figure 1(c) plots the temperature dependence of $dc$ magnetic susceptibility $\chi$ and reciprocal $1/\chi$ of FeCl$_2$ with a magnetic field of 0.1 T applied along the $c$ axis. One can clearly see a large drop in $\chi$ below ~ 23.2 K (see inset) in both zero-field-cooled (ZFC) and field-cooled (FC) measurements, indicative of an antiferromagnetic transition. The Curie temperature $\theta_{CW}$ extracted from the Curie-Weiss fit above 220 K is found to be ~ 84 K, and the effective magnetic moment $\mu_{eff}$ is about 4.46 $\mu_B$ which is slightly smaller than the spin-only value of 4.90 $\mu_B$ for Fe$^{2+}$ in a $3d^6$ configuration. Although FeCl$_2$ is antiferromagnetic, the positive sign of $\theta_{CW}$ stems from the dominant ferromagnetic interactions within the $ab$ plane. The isothermal magnetization $M(H)$ measured at various of temperatures are shown in Fig. 1(d). Below $T_N$, the magnetization increases sharply when the magnetic field reaches a critical value of $B_c \sim 1.5\,T$, characteristic of spin-flip transition, above which the system becomes a polarized ferromagnet, which is consistent with previous reports [29-31]. The small $B_c$ of the transition indicates weak antiferromagnetic interlayer couplings, affirming the vdW nature of this system.

Figure 2(a) and the inset present the temperature dependence of specific heat $C_P$

of $FeCl_2$ measured at various magnetic fields. At zero magnetic field, an anomaly occurs at 23.5 K, signaling the magnetic phase transition. Note that there is a small, additional anomaly near 21 K, which is not seen in the $\chi(T)$ data discussed above and in the previous neutron diffraction studies [21,22]. Such an anomaly was observed in two samples grown in different batches. It is presumably associated with the collapse of the entire magnon branches into a continuum scattering between 21 K (0.9 $T_N$) and $T_N$ as revealed by the inelastic neutron scattering studies [23]. In the presence of magnetic field, the anomaly associated with antiferromagnetic phase transitions is suppressed, while the magnitude of $C_P$ increases above $T_N$ compared to the values obtained at zero field. As an attempt to extract the magnetic heat capacity $C_{mag}$, using a Debye equation we fitted the $C_P(T)$ data measured at zero field between 2 K and 100 K excluding the 7 K – 80 K range. The thus-obtained the phonon contribution to the heat capacity, shown as the grey curve in Fig. 2(a), was then subtracted from $C_P$ data measured at various magnetic fields. The extracted $C_{mag}(T)$ curves are plotted in Fig. 2(b) where broad peaks around 35 K (above $T_N$) are observed whose magnitude increases in the presence of magnetic field up to 9 T. This feature suggests the onset of ferromagnetic correlation within the triangular planes which precedes the three-dimensional long-range antiferromagnetic order at $T_N$ due to the weak antiferromagnetic interlayer coupling [23]. In the presence of magnetic field, the associated Zeeman energy competes with the thermal energy and antiferromagnetic interlayer coupling. This may be responsible for enhancing the spin correlation between the triangular planes and thus leading to an increase in heat capacity above $T_N$. Note that the magnitude of heat capacity can be off by a factor of two due to the small mass (~ 0.3 mg) of the sample measured that is close to the uncertainty of the scale.

Figure 3(a) illustrates the schematic of the experimental setup for thermal transport measurements. Three temperature sensors and one heater were used. The heat current ($J_Q$) was applied in the *ab*-plane and the sign convention of $\kappa_{xx}$ and $\kappa_{xy}$ was determined by $\Delta T_{xx} = T_2 - T_1$ (in the direction of $J_Q$) and $\Delta T_{yx} = T_3 - T_1$ (in the direction of $B \times J_Q$), respectively (See Fig. S1). The magnetic field was applied along the *c*-axis. Figure 3(b) shows the temperature dependence of the longitudinal thermal conductivity $\kappa_{xx}(T)$ measured at various magnetic fields. In insulating magnetic solids, the thermal conduction is carried by magnons and phonons, thus $\kappa_{xx} = \kappa_{xx}^m + \kappa_{xx}^p$ where $\kappa_{xx}^m$ and $\kappa_{xx}^p$ represent the magnon and phonon contribution to the thermal conductivity, respectively. In the absence of magnetic field, no anomalous feature in $\kappa_{xx}(T)$ is observed at $T_N$ as would be expected from an exchange-dominated magnetostriction. Instead, an anomalous dip in $\kappa_{xx}(T)$ between two broad peaks was observed below $T_N$, consistent with early reports [32-34], which implies phonon scattering by magnons. Below the low temperature peak around 5.6 K, $\kappa_{xx}$ is dominated by $\kappa_{xx}^p$ because of the small population of magnons due to the single-ion anisotropy gap of ~ 2.2 meV at the zone center of magnetic excitations [23]. In the intermediate temperature region, the population of thermally populated magnons increases, which scatter phonons via the strong magnon-phonon coupling and result in a decrease in $\kappa_{xx}^p$. Such a decrease in $\kappa_{xx}^p$ even overcomes the thermal contribution of magnons, leading to the reduction of the total $\kappa_{xx}$. As the temperature further increases and approaches $T_N$, magnons become less coherent with a decrease in magnon lifetime as inferred from inelastic neutron scattering studies [23]. Thus, the phonon scattering by magnons decreases, which then increases the total $\kappa_{xx}$ again until above 40 K at which point the Umklapp scattering dominates leading to a decrease in $\kappa_{xx}$. A similar argument was made in Ref.[30] to account for the zero-field $\kappa_{xx}(T)$ features. With an applied magnetic field larger than $B_c$, FeCl$_2$ becomes a polarized ferromagnet and the magnon gap value increases with increasing magnetic field, giving rise to the reduction of magnon population. As a result, magnon-phonon scattering is suppressed and the phonon mean free path increases, resulting in an increase in $\kappa_{xx}^p$ and thus total $\kappa_{xx}$ in the low temperature region. On the other hand, as suggested by the specific heat measurement shown in Fig. 2 and discussed above, ferromagnetic correlation within the triangular planes sets in above $T_N$. An application of magnetic field enhances three-dimensional ferromagnetic correlation, which leads to more coherent magnons that scatter phonons and consequently decrease the total $\kappa_{xx}$ compared to $\kappa_{xx}$ measured at zero field. Future inelastic neutron scattering studies of FeCl$_2$ in the presence of magnetic field is desirable to shed light on this.

Figures 3(c,d) show the magneto-thermal conductivity, which is defined as $(\kappa_{xx}(B) - \kappa_{xx}(0)) \times 100\%/\kappa_{xx}(0)$, as a function of magnetic field measured at various temperatures. One can see that FeCl$_2$ exhibits very large magneto-thermal conductivity, reaching over 700% around 10 K at 9 T, again suggesting strong phonon scattering by magnons which is suppressed in the presence of a high field as discussed above. The magnetic field dependence of magnon bands of FeCl$_2$ is illustrated in Fig.

S10 in the Supplemental Materials [25]. Below $T_N$, at zero magnetic field FeCl$_2$ has two magnon bands that are degenerate. An application of a magnetic field smaller than $B_c$, when FeCl$_2$ remains antiferromagnetic, breaks the degeneracy of the magnon bands by shifting down one magnon branch in energy while pushing up the other magnon branch due to the Zeeman interaction. As a result, the phonon scattering by magnons is slightly enhanced and the resultant $\kappa_{xx}^p$ and the total $\kappa_{xx}$ decrease, as shown in Fig. 3(c). Above $B_c$, as discussed above, FeCl$_2$ becomes ferromagnetic via the spin-flip transition and the magnon gap increases with magnetic field. Thus, the phonon scattering by magnons is suppressed due to the decrease in the magnon population, leading to positive magneto-thermal conductivity. As shown in Fig. 3(d), similar features in magneto-thermal conductivity are observed above $T_N$, although with much smaller magnitude, which is presumably ascribed to the (para)magnons associated with ferromagnetic correlation with the triangular planes that diminishes around 80 K.

Next, we present the thermal Hall effect observed in FeCl$_2$. To unambiguously validate the intrinsic nature of the measured thermal Hall signal, we have measured the thermal Hall signal, sweeping the magnetic field from -9T to 9T and back and following the procedure reported in Ref. [3,35] to examine its hysteretic behavior. In addition, measurements were also performed using different heating powers and on different samples to ensure the intrinsic nature of the THE and its reproducibility. Detailed procedures for extracting the thermal Hall signal and the thermal transport data measured on another sample are described in the Supplemental Materials (Fig. S2 – S9) [25].

Figure 4(a) presents the field dependent thermal Hall conductivity $\kappa_{xy}$ measured at various temperatures. A large thermal Hall effect is clearly observed. Below $T_N$, $\kappa_{xy}$ increases with increasing magnetic field and then sharply switches sign near 1.5 T which coincides with the spin-flip transition at $B_c$, as also shown in Fig. S7. Upon increasing the temperature, the magnitude of $\kappa_{xy}$ just above $B_c$ increases, reaching 0.05 W m$^{-1}$ K$^{-1}$ with a large thermal Hall angle ($\theta_{th} = \kappa_{xy}/\kappa_{xx}$) of 0.008 at $T \sim 18$ K (shown in Fig. S6). After the sign change, the magnitude of both $\kappa_{xy}$ and $\theta_{th}$ decrease when further increasing the magnetic field. These features exclude the possibility that the observed thermal Hall signal in FeCl$_2$ is (1) purely driven by phonons as reported in nonmagnetic SrTiO$_3$ [36] since one would expect $\kappa_{xy}$ monotonically increases with magnetic field; (2) driven by phonon THE induced by the internal magnetic field of magnetization since no sign change in $\kappa_{xy}$ would be expected. In addition, as will be discussed next, the observed thermal Hall signal cannot be ascribed to topological magnons either, since the magnon bands are topologically trivial. Instead, the observed THE in FeCl$_2$ can be associated with the magnon-phonon hybridization [24].

To theoretically study the THE in FeCl$_2$ below $T_N$, we first consider the monolayer triangular lattice model and take into account the magnon-phonon coupling. The associated Hamiltonian, parameters, and detailed calculations are discussed in Section V in the Supplemental Materials [25]. As shown in Fig. S11, with the inclusion of the phonon bands and the introduction of magnon-phonon coupling, a gap opens between the hybridized bands. These hybridized bands exhibit nontrivial topological Berry curvature, from which $\kappa_{xy}$ of the ferromagnetic monolayer triangular lattice may be calculated (Fig. S12 with additional details found in the Supplementary Materials [25]).

The THE driven by magnons [37] and magnon polarons [38] has been very recently studied theoretically in honeycomb antiferromagnets, and shown to be sensitively dependent on the external magnetic field which can induce both topological and magnetic phase transitions. Such a dependence on the magnetic field is similarly expected to influence the profile of the thermal Hall signal in antiferromagnetically-coupled FeCl$_2$ bilayers. To incorporate the effects of the weak antiferromagnetic interlayer interaction in FeCl$_2$, we employ an effective bilayer approach where, in the antiferromagnetic phase, the total thermal Hall conductivity $\kappa_{xy}$ is given by $\kappa_{xy} = (\kappa_{xy}^\uparrow + \kappa_{xy}^\downarrow)/2$ with $\kappa_{xy}^{\uparrow(\downarrow)}$ being the thermal Hall conductivity calculated from $+(-)\hat{B}$-polarized magnons in the monolayer calculation. In the presence of a weak external field, this approach assumes decoupled magnon modes of opposing polarization in the bilayers, an assumption that is predicated on the weak interlayer coupling and strong single-ion anisotropy exhibited by FeCl$_2$. Beyond the critical field $B_c$, the system becomes a polarized ferromagnet where $\kappa_{xy}$ is simply given by $\kappa_{xy}^\uparrow$.

The calculated thermal Hall conductivity due to magnon-phonon coupling in bilayer FeCl$_2$ is shown in Fig. 4(b) as a function of magnetic field for various temperatures. The critical field $B_c$ at which the induced ferromagnetic phase begins is approximated from the experimental data. A striking similarity with experimental results is observed in the field dependence of the calculated bilayer $\kappa_{xy}$, especially regarding the sign change at $B_c$. As illustrated in Fig. S10, at zero field, the ↑ and ↓ magnon bands are degenerate and, since these systems are related to one another through a $\pi$-rotation about the $x$-axis, their thermal Hall conductivities cancel each other exactly. Introduction of a small external field breaks the degeneracy of the ↑/↓ magnon bands, leading to suppression of the ↑ magnon population and allowing the contribution from the ↓ magnon band to dominate. In the induced ferromagnetic phase right above $B_c$, however, only the ↑ magnon band with an enhanced population remains to determine $\kappa_{xy}$, leading to a sign change in the overall thermal Hall signal with an increase in the magnitude of $\kappa_{xy}$. As the external field continues to increase in magnitude, $\kappa_{xy}$ is further suppressed as the magnon band shifts higher in energy, increasing the energy of magnon-phonon hybridization and reducing the magnon population. These theoretical results qualitatively support the notion that the THE is partially contributed by magnon-phonon hybridization in FeCl$_2$, and the characteristic sign change in $\kappa_{xy}$ is indicative of the transition from the layered antiferromagnetic phase to the induced ferromagnetic phase. We note that there is a quantitative discrepancy in $\kappa_{xy}$ between the experimental data and the theoretical calculation, particularly at higher temperatures. Such a discrepancy may be attributed to features that are not included in the model (e.g., coupling with additional phonon modes, higher-order fluctuation terms, etc.). In addition, our theoretical results do not discount the possibility of contributions to the THE from other mechanisms such as magnon-phonon scattering. Unfortunately, the contribution from magnon-phonon scattering to the THE is seldom discussed in the literature, which warrants future studies. While the extent to which magnon-phonon hybridization contributes to the observed thermal Hall signal cannot be established quantitatively with the present methods, we believe that the theoretical calculation presented here serves as a proof-of-concept model providing a qualitative description of the essential physics underlying the shape of the $\kappa_{xy}$ signal.

In conclusion, we report the observation of large magnetothermal conductivity and THE in a vdW antiferromagnet $FeCl_2$, elucidating strong magnon-phonon coupling in $FeCl_2$. The thermal Hall signal changes the sign upon the spin-flip transition from the antiferromagnetic state to the polarized ferromagnetic state. The observed THE is in qualitative agreement with the THE arising from the Berry curvature at the anticrossing points of hybridized magnon and phonon modes, suggesting that the THE at least is partially contributed from magnon-polarons. Nevertheless, the discrepancy in the magnitude of THE between the calculation result and experimental data suggests that other mechanisms can also play a role, which calls for further theoretical studies.

**Acknowledgements**

H.Z., M.S., and X.K. acknowledge the financial support by the U.S. Department of Energy, Office of Science, Office of Basic Energy Sciences, Materials Sciences and Engineering Division under DE-SC0019259. X.K. also acknowledges the financial support by the National Science Foundation (DMR-2219046). X.K. appreciates the insightful discussion with Prof. S.D. Mahanti. P.Z. acknowledges the financial support by the National Science Foundation (DMR-2112691). C.C. and D.X. are supported by AFOSR MURI 2D MAGIC (FA9550-19-1-0390). C.X. is partially supported by the Start-up funds at Michigan State University.

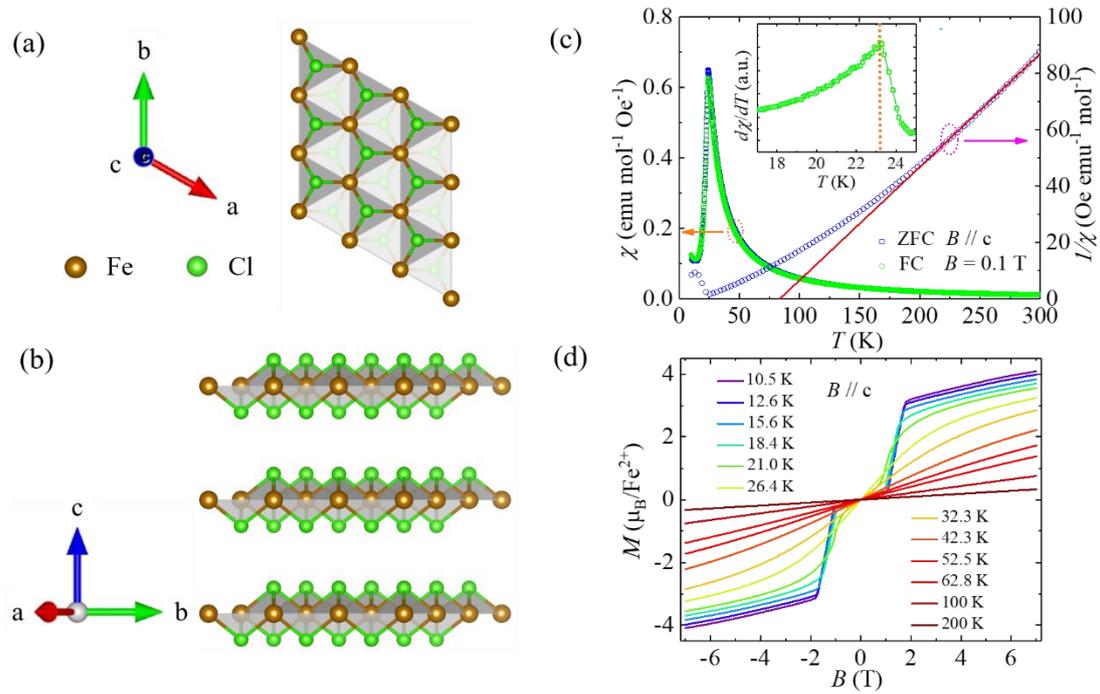

**Figure 1. Crystal structure and magnetization of FeCl$_2$**. (a, b) Schematic crystal structure of FeCl$_2$ viewed along two different perspectives. (c) Temperature dependence of *dc* magnetic susceptibility $\chi$ with 0.1 T magnetic field applied along the *c* axis. The red solid linear line is the Curie-Weiss fitting and the inset is an expanded view of $d\chi/dT$ at low temperatures. (d) Isothermal magnetization curves measured at various temperatures.

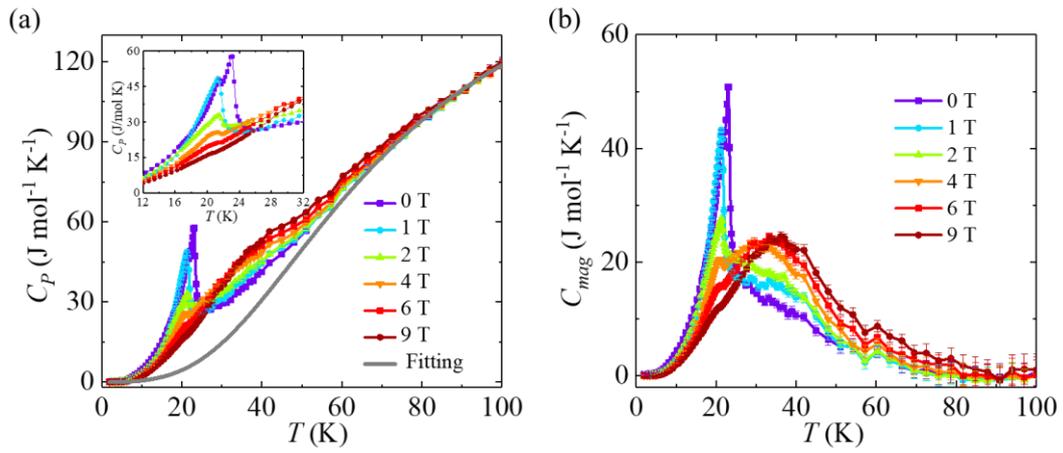

**Figure 2. Specific heat of FeCl$_2$**. (a) Temperature dependence of specific heat $C_p$ measured at various magnetic fields. The grey curve is a fitting based on a Debye equation for the phonon contribution. The inset shows an expanded view. (b) Temperature dependence of the extracted magnetic specific heat $C_{mag}$.

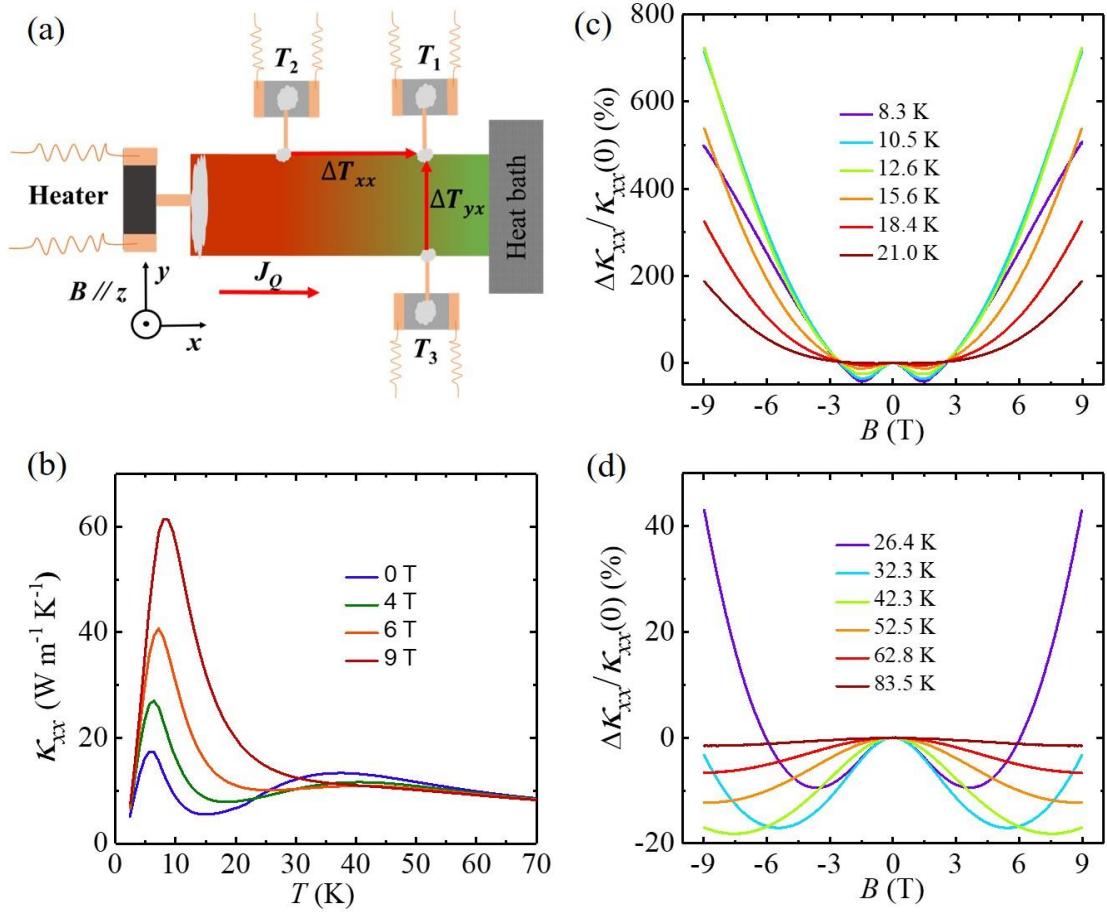

**Figure 3. Thermal conductivity of FeCl$_2$**. (a) A schematic of the experimental setup for thermal transport measurements. (b) Thermal conductivity $\kappa_{xx}$ as a function of temperature measured with various magnetic fields. (c) and (d) Magneto-thermal conductivity measured at various temperatures. The magnetic field is applied along the $c$ axis.

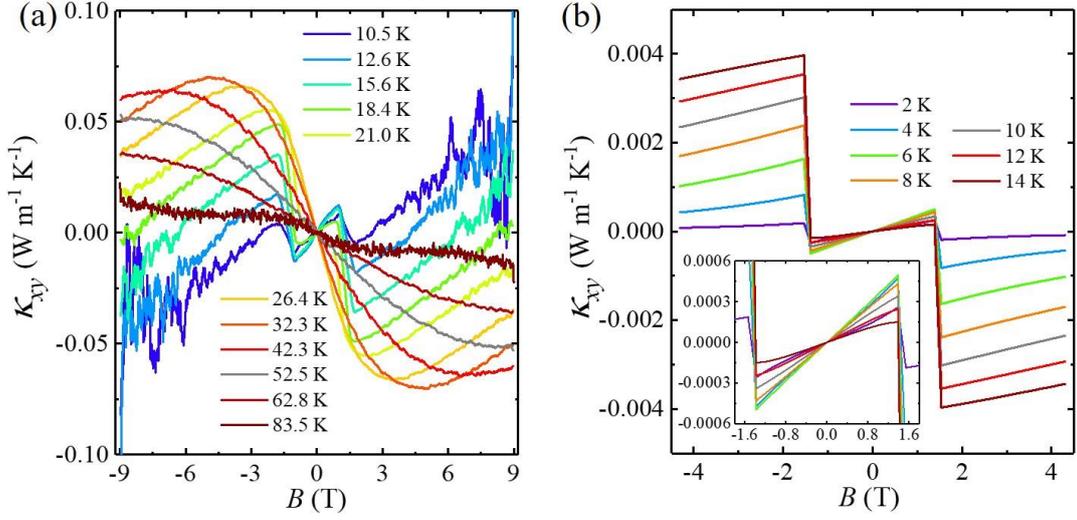

**Figure 4. Thermal Hall conductivity of FeCl$_2$**. (a) Magnetic field dependence of thermal hall conductivity $\kappa_{xy}$ measured at various temperatures. (b) The calculated bilayer thermal Hall conductivity with magnon-phonon coupling constant $g = 1.0$ (see Supplemental Materials [25] for details) assuming a transition to an induced ferromagnetic phase at the critical field of $B_c = 1.5$ T. Below $B_c$, $\kappa_{xy} = (\kappa_{xy}^{\uparrow} + \kappa_{xy}^{\downarrow})/2$ and above $B_c$, $\kappa_{xy} = \kappa_{xy}^{\uparrow}$. Inset: an expanded view of the temperature dependence in the antiferromagnetic weak-field region.